\documentclass[prb,superscriptaddress,showpacs,aps,twocolumn]{revtex4-1}
\usepackage{amsmath}
\usepackage{graphicx}
\usepackage{dcolumn}
\usepackage{bm}
\usepackage{epsfig}

\newcommand{\Rvec}{{\bf R}}
\newcommand{\qvec}{{\bf q}}
\newcommand{\rvec}{{\bf r}}

\newcommand{\yvec}{{\bf y}}
\newcommand{\Qvec}{{\bf Q}}
\newcommand{\beq}{\begin{equation}}
\newcommand{\eeq}{\end{equation}}
\newcommand{\bea}{\begin{eqnarray}}
\newcommand{\eea}{\end{eqnarray}}
\begin{document}
\title{Iterative backflow renormalization procedure for  many-body ground state wave functions of strongly interacting normal Fermi liquids}

\author{Michele Taddei}
\affiliation{Dipartimento di Fisica, Sapienza Universit\`a di Roma, Piazzale A. Moro 2, I-00185, Roma, Italy}
\author{Michele Ruggeri}
\affiliation{DEMOCRITOS National Simulation Center, Istituto Officina dei Materiali del CNR and SISSA, 
Via Bonomea 265, I-34136 Trieste, Italy}
\author{Saverio Moroni}
\affiliation{DEMOCRITOS National Simulation Center, Istituto Officina dei Materiali del CNR and SISSA, 
Via Bonomea 265, I-34136 Trieste, Italy}
\author{Markus Holzmann}
\affiliation{LPTMC, UMR 7600 of CNRS,  Universit{\'e} Pierre et Marie Curie, Paris,
France}
\affiliation{LPMMC, UMR 5493 of CNRS, Universit{\'e} Grenoble Alpes, F-38100 Grenoble
France}
\affiliation{Institut Laue Langevin, BP 156, F-38042 Grenoble Cedex 9, France}

\date{\today}
\begin{abstract}
We show how a ground state trial wavefunction of a Fermi liquid can 
be systematically improved introducing a sequence of renormalized 
coordinates through an iterative backflow transformation. 
We apply this scheme to calculate the ground state energy of liquid 
$^3$He in two dimensions at freezing density using variational
and fixed-node diffusion Monte Carlo.
Comparing with exact transient estimate results for systems with small
number of particles, we find that variance extrapolations provide accurate
results for the true ground state together with stringent lower bounds. For
larger systems these bounds can in turn be used to quantify the systematic 
bias of fixed-node calculations. These wave functions are size consistent and the
scaling of their computational complexity with the number of particles is the same as 
for standard backflow wave functions.
\end{abstract}
\pacs{PACS: }
\maketitle

\section{Introduction}
To overcome the fermion-sign problem, many fermion Quantum Monte Carlo (QMC)
calculations rely on the fixed-node (FN) approximation where the
nodes of a trial wavefunction, $\psi_T$, are imposed as a boundary
condition on the many-body Schr{\"o}dinger equation which can then be
solved by projector  Monte Carlo methods \cite{Reynolds82}.
Since the nodal surfaces of the exact ground state wavefunction
are in general unknown, the energies of FN calculations do not converge
to the exact ground state energy but remain above them
by an unknown amount. Although methods which do not rely on the FN approximation
have been developed\cite{RN,Carleo,Nava,Alavi,shadows}, they are in general limited to small systems as
their computational cost grows exponentially with system size. Therefore, FN-QMC calculations
still provide the most accurate values of ground state properties of extended fermion systems. 

Modification of the nodes of a many-fermion wave function to explicitly include correlations
remains a formidable task. Slater determinants based on
Backflow (BF) coordinates present one possibility
\cite{Panda73,Schmidt79,Manousakis83,Vitiello97}, and  backflow wave functions have been
routinely used over the last years in QMC calculations
of the electron-gas
\cite{Kwon93,Kwon98,BF} and liquid $^3$He
\cite{Schmidt81,Panoff89,Casulleras00,Moroni95,Zong03}. Generalization of the backflow 
wave function to include  three body correlations
was shown to  be necessary to stabilize the unpolarized phase of liquid $^3$He 
against spin-polarization  \cite{BF3}.

Here we propose new correlated trial wave functions based on iterative backflow transformations
and use them to study liquid $^3$He in two dimensions.
We show that this new class of trial wave functions systematically lower the energy and its variance.
Our results illustrate the possibility to extrapolate variational (VMC) and FN diffusion (DMC) Monte Carlo
calculations to zero variance to approach very closely the {\em exact} ground state energy.
Since their evaluation remains of similar
complexity and scaling with increasing system size as the usual backflow wave function,
their use is not limited to small systems. 
We explicitly demonstrate
the size consistency of our new trial wave functions and discuss the possibility to obtain 
lower bounds to the ground state energy.

\section{Iterated renormalization of wave function}
\label{heuristic_derivation}
Let us start by considering the standard Slater-Jastrow type trial wave function 
with backflow,
\beq
\Psi_T^{(0)} = \det \phi_k (\qvec_i [ \Rvec] ) e^{-U[\Rvec]}.
\label{SJ}
\eeq
Antisymmetry is 
ensured by the Slater determinant of single particle orbitals, $\phi_k(\rvec)$, $k=1,\dots, N$,
where, instead of the bare coordinates $\rvec_i$, $i=1,\dots, N$, 
 many-body backflow coordinates, $\qvec_i$,  are used 
as arguments.
Both backflow coordinates, $\Qvec=(\qvec_1,\dots,\qvec_N)$,
and the symmetric Jastrow potential, $U$, depend explicitly
on all coordinates, $\Rvec=(\rvec_1, \rvec_2,\dots, \rvec_N)$, as indicated. In the standard form,
$U=\sum_{i<j} u(r_{ij})+\sum_i{\bf G}_i(\Rvec)\cdot{\bf G}_i(\Rvec)$ with
${\bf G}_i=\sum_j(\rvec_i-\rvec_j)\xi(r_{ij})$, $\qvec_i=\rvec_i + \sum_j (\rvec_i-\rvec_j) \eta(r_{ij})$,
and $r_{ij}=|\rvec_i-\rvec_j|$.
The radial functions $u$, $\xi$ and $\eta$
can be parametrized and optimized by minimization of the variational energy. Generalizations to include higher correlations 
into both backflow and Jastrow potentials are possible \cite{BF3} but will not be considered here.

Once the backflow and Jastrow potentials have been determined, 
different occupations of the orbitals inside the Slater determinant of  
Eq.~(\ref{SJ}) can be used to approximate also low-lying excited states 
of the systems \cite{McMillan65}, in close
analogy to Landau's Fermi liquid description. As in the correlated basis 
functions approach \cite{Feenberg,Krotscheck}, let us consider the effective 
Hamiltonian within these non-orthogonal basis states.
For a Fermi liquid, we expect non-diagonal matrix elements of the effective 
Hamiltonian to be strongly suppressed compared to those of the bare 
plane-wave states. However, instead of diagonalizing the effective 
Hamiltonian, let us search again for a trial wave function to represent 
the ground state of the effective Hamiltonian. Assuming a smoothly varying 
effective interaction, we may again consider to represent it
as a backflow wave function, $\Psi_T^{(1)}$. However, this time, 
the new backflow coordinates, $\qvec_i^{(1)}$,
and the new Jastrow potential, $U^{(1)}$, are
built upon the old backflow coordinates, $\Qvec^{(1)}[\Qvec^{(0)}]$, 
and $U^{(1)}[\Qvec^{(0)}]$ with $\Qvec^{(0)}\equiv \Qvec$. 
Thus we are naturally led to an iterative renormalization procedure
\beq
\Psi_T^{(\alpha)} \rightarrow \Psi_T^{(\alpha+1)} 
= \det
\phi_k(\qvec_i^{(\alpha+1)}) e^{- U^{(\alpha+1)}}
\eeq
with a renormalized Jastrow potential
\bea
U^{(\alpha)}  & = & \sum_{\beta \le \alpha} \Big[ \sum_{i<j} 
u^{(\beta)}\left(q_{ij}^{(\beta-1)}\right) 
\nonumber \\
              & + & \sum_i{\bf G}_i^{(\beta)}(\Qvec^{\beta-1})
                     \cdot{\bf G}_i^{(\beta)}(\Qvec^{\beta-1}) \Big]  
              \label{Jastrowrenorm}
\eea
and renormalized backflow coordinates
\bea
\label{Backflowrenorm}
\qvec_i^{(\alpha)} &=& \rvec_i+\sum_{\beta \le \alpha} \yvec^{(\beta)}_i \\
\yvec_i^{(\alpha)}[\Qvec^{(\alpha-1)}] & =&  
\sum_{j\ne i}
\left(\qvec_i^{(\alpha-1)}-\qvec_j^{(\alpha-1)}\right) 
\eta^{(\alpha)}\left(q_{ij}^{(\alpha-1)}\right)
\nonumber
\eea
(in Eqs. (\ref{Jastrowrenorm}) and (\ref{Backflowrenorm}), $\Qvec^{(-1)}$ stands
for $\Rvec$).
At each iteration new potentials parametrizing the additional Jastrow and 
backflow functions are introduced,
and all the potentials $u^{(\beta)}$, $\xi^{(\beta)}$ and $\eta^{(\beta)}$, 
with $\beta \leq\alpha$, have to be optimized.
 
\begin{table*}
 \begin{tabular}{|l|l|r|c|l|l|r|c|l|l|r|c|l|} \hline
  & \multicolumn{4}{c|}{N=26, $\zeta=0$} & \multicolumn{4}{c|}{N=58, $\zeta=0$} & \multicolumn{4}{c|}{N=29, $\zeta=1$}  \\
  \hline
           & $E_{T}/N$ & $\sigma^2/N$ & $\Delta$ & $E_{DMC}/N$ & $E_{T}/N$ & $\sigma^2/N$ & $\Delta$ & $E_{DMC}/N$ & $E_{T}/N$ & $\sigma^2/N$ & $\Delta$ & $E_{DMC}/N$\\
  \hline
PW         & 3.011(1) & 28.29 &       & 2.419(2)   &   2.900(1) & 28.07 &       & 2.373(2) & 2.5831(6) & 7.51  &       &  2.402(1)  \\
BF$^{(0)}$ & 2.688(1) & 13.05 & 0.323 & 2.353(2)   &   2.584(1) & 13.34 & 0.316 & 2.283(2) & 2.5133(5) & 5.34  & 0.070 &  2.4005(6) \\
BF$^{(1)}$ & 2.471(1) & 4.58  & 0.540 & 2.336(2)   &   2.356(2) & 4.93  & 0.544 &          & 2.4383(3) & 2.20  & 0.145 &  2.3918(5) \\
BF$^{(2)}$ &2.4258(8) & 2.86  & 0.585 & 2.3284(9)  &   2.313(2) & 3.25  & 0.587 &          & 2.4193(3) & 1.54  & 0.164 &  2.3877(4) \\
BF$^{(3)}$ &2.4049(9) & 2.47  & 0.606 & 2.3223(4)  &   2.297(2) & 2.67  & 0.603 &          & 2.4136(2) & 1.36  & 0.170 &  2.387(1)  \\
BF$^{(4)}$ & 2.400(1) & 2.29  & 0.611 & 2.323(1)   &   2.292(1) & 2.49  & 0.608 & 2.232(1) & 2.4109(7) & 1.25  & 0.173 &  2.3869(5) \\
    \hline 
VMC$_{ext}$ & \multicolumn{4}{ c|}{2.338(5)}        & \multicolumn{4}{ c|}{2.217(2)}  &  \multicolumn{4}{ c|}{2.384(6)}   \\
DMC$_{ext}$ & \multicolumn{4}{ c|}{2.317(3)}        & \multicolumn{4}{ c|}{2.216(3)}  &  \multicolumn{4}{ c|}{2.379(1)}   \\
LB$_{ext}$  & \multicolumn{4}{ c|}{2.275(14)}       & \multicolumn{4}{ c|}{2.149(12)} &  \multicolumn{4}{ c|}{2.390(26)}  \\
    \hline 
TE         & \multicolumn{4}{ c|}{2.307(7)}        & \multicolumn{4}{ c|}{        }  &  \multicolumn{4}{ c|}{2.375(3)}   \\
 \hline
 \end{tabular}
\caption{
Ground-state energy per particle, in K,
of liquid $^3$He in two dimensions at $\rho=0.060$\AA$^{-2}$, 
obtained with Variational ($E_{T}/N$) and fixed node Diffusion Monte Carlo ($E_{DMC}/N$) using different types of 
trial wave functions: Slater-Jastrow wave function without backflow (PW), and with $\alpha$-times iterated 
backflow (BF$^{(\alpha)}$).
$\zeta$ is the spin polarization and $N$ is the number of particles.
$\Delta$ is the gain in VMC energy per particle relative to the PW value, and $\sigma^2$ is 
the variance of the VMC total energy.
TE indicates unbiased results calculated with the transient estimate method of Ref. \onlinecite{Carleo}. 
VMC$_{ext}$, DMC$_{ext}$ and LB$_{ext}$ are the extrapolations to zero variance of $E_T/N$, $E_{DMC}/N$ and
of the lower bound $(E_T-\sqrt{\sigma^2})/N$, respectively.
Statistical uncertainties on the last digit(s) are given in parentheses.
All values are given for periodic boundary conditions ($\Gamma$ point) without tail corrections\cite{Aziz}. 
}
\label{energies}
\end{table*}

In the appendix we show how  the evaluation of the renormalized wave functions
and their derivatives needed to calculate the local energy 
can be efficiently implemented with a number of operations proportional to $N^3$.
Thus, the overall cost of calculation is not dramatically altered compared to the usual 
(zeroth order) backflow wave function. For a system of $N=26$ particles 
we find that the CPU time to move all the particles and calculate the local energy with iterated backflow of order 1 to 4 
is a factor 5, 9, 13, 17 larger than that of the zeroth order, respectively; 
furthermore, for $N=58$, fourth order backflow takes 12.5 times longer than for $N=26$, close to 
the $N^3$ scaling. The corresponding figures for the efficiency of the calculation of the energy are even 
more favorable, because the variance is lower for improved wave functions.

\section{Two dimensional liquid $^3$He at freezing density}
In order to illustrate the accuracy of the renormalization procedure to describe the ground state wave function
of highly correlated Fermi liquids, we perform calculations for the ground state energy of
liquid $^3$He in two dimensions at a density
$\rho=0.060$\AA$^{-2}$, near freezing\cite{Nava}. We compare VMC and fixed node DMC energies to exact results
obtained by the nominally exact transient estimate (TE) method of Ref. \onlinecite{Carleo},
for systems of $N=26$ ($N=29$) unpolarized (polarized) $^3$He atoms 
interacting
with the HFDHE2 potential \cite{Aziz}. Furthermore we test the size-consistency of our trial functions,
comparing the gain in variational energy obtained by the renormalization procedure for the unpolarized system
at two different sizes, $N=26$ and $N=58$. The results are collected in Table \ref{energies}.

\begin{figure}
\includegraphics[width=\textwidth]{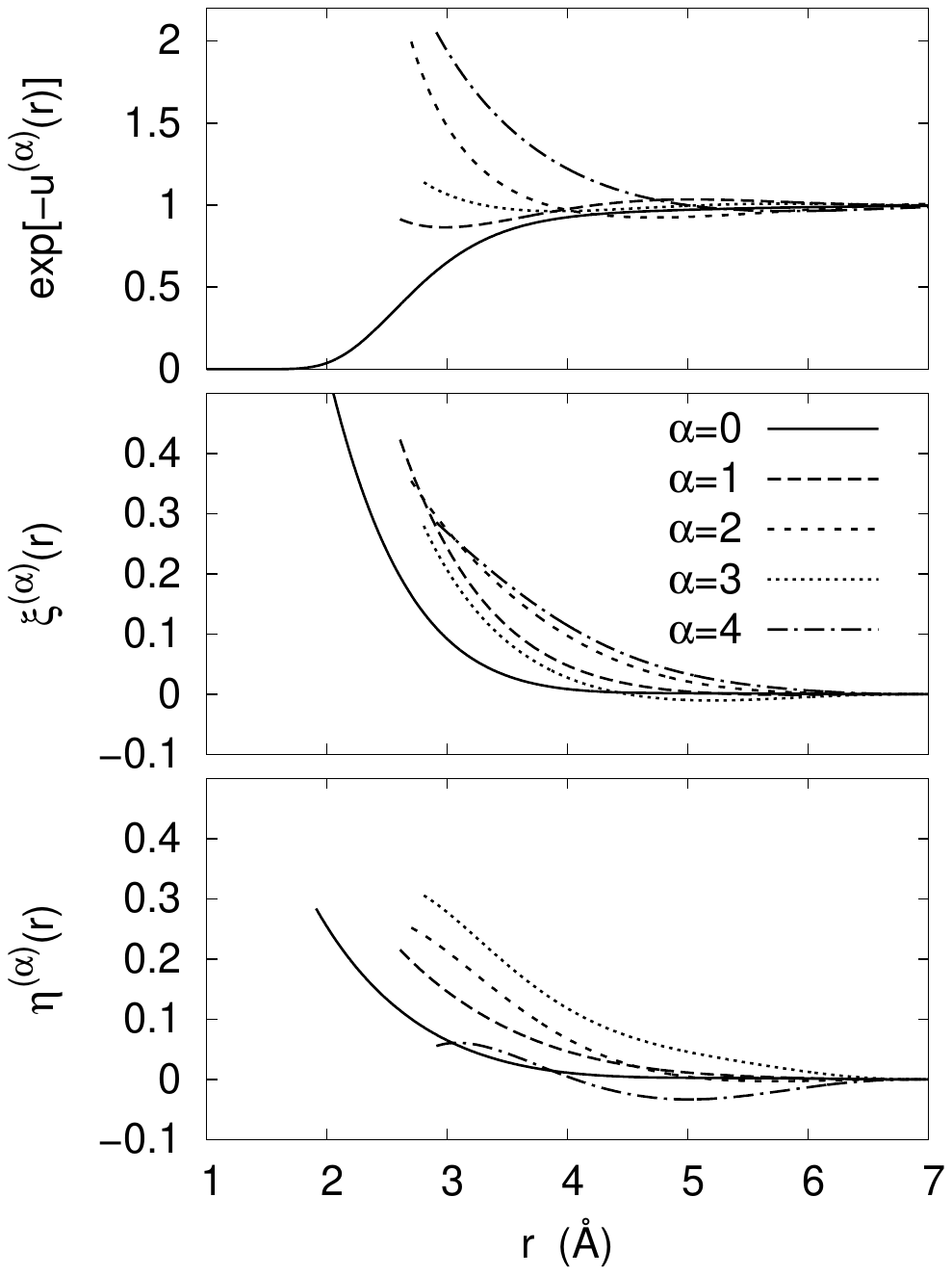}
  \caption{Optimized potentials of the trial function $\Psi_T^{(4)}$
for $N=26$, $\zeta=0$. The lines are broken where 
the pair distribution functions of the relevant (quasi)coordinates
become negligibly small, $g(r)\lesssim 10^{-3}$ (see Fig.~\ref{gr}).
}
\label{alll}
\end{figure}

\begin{figure}
\includegraphics[width=0.5\textwidth]{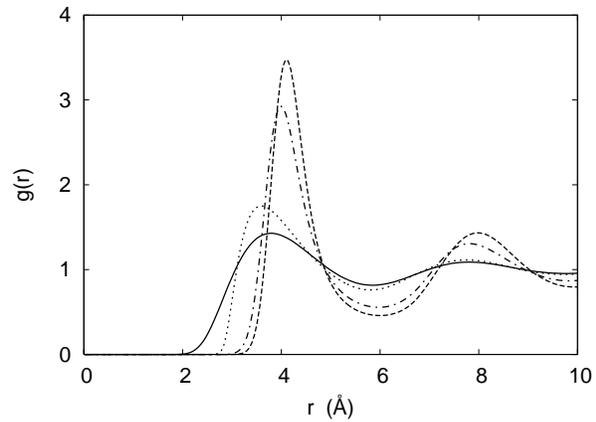}
  \caption{Pair correlation functions calculated in a VMC simulation with the BF$^{(4)}$ trial function using
the bare coordinates $\{{\bf r}_i\}$ (solid line), and the
renormalized backflow coordinates  $\{{\bf q}_i^{(0)}\}$, $\{{\bf q}_i^{(2)}\}$, and $\{{\bf q}_i^{(4)}\}$ (dotted, dash-dotted
and dashed lines, respectively). The statistical noise reaches its maximum value $\sim 0.003$ at the highest peak. 
}
\label{gr}
\end{figure}

Every iteration introduces three new potentials (for backflow, two- and three-body Jastrow function), each of which,
generically indicated here as $f(r)$, is parametrized in the form
\begin{equation}
f(r)=\begin{cases} (r_C-r)^3\left[\sum_{n=1}^5a_nr^{n-1}+a_6/r^{a_7}\right] &\mbox{if } r<r_C\\ 0 &\mbox{if } r\geq r_C. \end{cases}
\label{potentials}
\end{equation}
For the backflow and three-body Jastrow potentials we set $r_C=7$\AA~ and drop the McMillan term ($a_6=0$), while for the two-body Jastrow 
potential we choose a cutoff value $r_C$ close to half the side of the simulation box. In Fig.~\ref{alll} we show the optimized potentials $u^{(\alpha)}$, $\xi^{(\alpha)}$ and $\eta^{(\alpha)}$ of the $\Psi_T^{(4)}$ wave function for a 
system with $N=26$ and $\zeta=0$. 
The backflow coordinate transformations across different iterations 
implicitly build up many-body correlations at all orders, so that 
eventually not all of the optimized potentials have an obvious physical 
interpretation: 
for instance the pair distribution functions $g$ of the bare coordinates
and of the renormalized coordinates at subsequent iteration levels,
shown in Fig.~\ref{gr}, feature increasingly wide correlation holes 
and high peaks for increasing level, despite the two-body potentials 
$u^{(\alpha)}$ turning from repulsive for $\alpha=0$ to attractive for $\alpha=4$.
Note that all the $g$'s feature the structure of simpèle liquids 
(albeit with increasingly classical character), which supports the
heuristic derivation given in Section~\ref{heuristic_derivation}: 
each iteration essentially renormalizes the Slater Jastrow wave 
function without qualitative changes. 

With the choice of Eq.~(\ref{potentials}), the renormalization
procedure requires 17 variational parameters per level, and the 
corresponding optimization procedure (carried out by correlated 
sampling \cite{reweighting} in this work) becomes rather demanding. 
Therefore we have tried two simpler
iterative schemes, one in which no renormalized Jastrow is present, and one in which only the new potentials added at the $\alpha$-th
iteration are optimized, leaving the other unchanged from previous iterations. However, these simpler options
lead to higher values in energy, both for VMC and DMC.
We have also considered an improved wave function with different backflow potentials for parallel and antiparallel spins. The gain
in energy is $\sim 10$ mK in VMC, but hardly visible in DMC ($\lesssim 1$ mK) beyond the second backflow iteration.
Finally, we have tested the accuracy of using potentials optimized for $N=26$ to perform simulations with $N=58$ particles:
at the fourth backflow iteration, the DMC energy is higher by a non negligible amount, 5$\pm$2 mK.
All these results are listed in Table \ref{energies_bis}.

\begin{table*}
 \begin{tabular}{|l|l|l|l|l|l|l|l|l|l|l|l|l|} \hline
  & \multicolumn{8}{c|}{$N=26$, $\zeta=0$} & \multicolumn{4}{c|}{$N=58$, $\zeta=0$} \\
  \hline
           & \multicolumn{4}{c|}{$E_T/N$}              & \multicolumn{4}{c|}{$E_{DMC}/N$}              & \multicolumn{2}{c|}{$E_T/N$} & \multicolumn{2}{c|}{$E_{DMC}/N$}  \\
  \hline
           &0         &I         &II        &III       &0            &I         &II        &III        &0         &IV        &0         &IV        \\
  \hline
PW         &          &          &          &          &             &          &          &           & 2.900(1) & 2.909(2) &          &          \\
BF$^{(0)}$ &          &          &          &          &             &          &          &           & 2.584(1) & 2.592(2) & 2.289(2) & 2.288(1) \\
BF$^{(1)}$ & 2.471(1) & 2.599(2) & 2.515(1) & 2.461(1) &   2.336(2)  & 2.337(2) & 2.337(2) &           &          &          &          &          \\
BF$^{(2)}$ &2.4258(8) & 2.585(2) & 2.480(1) & 2.413(1) &   2.3284(9) & 2.335(2) & 2.332(1) & 2.3256(9) &          &          &          &          \\
BF$^{(3)}$ &2.4049(9) & 2.584(2) & 2.472(1) & 2.398(1) &   2.3223(4) & 2.335(2) & 2.326(1) & 2.3215(4) &          &          &          &          \\
BF$^{(4)}$ & 2.400(1) & 2.580(2) & 2.470(1) & 2.390(2) &   2.323(1)  & 2.331(1) & 2.325(3) & 2.324(1)  & 2.292(1) & 2.298(2) & 2.232(1) & 2.237(1) \\
    \hline 
 \end{tabular}
\caption{Some of the energies of Table \ref{energies} compared to the corresponding 
values obtained with downgraded or upgraded wave functions. Entries 0: energies from Table \ref{energies}; 
entries I: downgraded wave functions with omitted Jastrow factors of the quasi-coordinates; 
entries II: downgraded wave functions with Jastrow and backflow
potentials from previous iterations not repotimized; entries III: upgraded wave functions with different
like-spin and unlike-spin backflow potentials; entries IV: downgraded wave functions for $N=58$ 
with Jastrow and backflow potentials optimized for $N=26$.
 }
\label{energies_bis}
\end{table*}

\section{Zero-variance extrapolation and lower bounds}
Our VMC and FN-DMC results for the energy expectation values
$E_X=\langle \Psi_X | H | \Psi_X \rangle$ of the different (normalized) wave
functions, provide strict upper bounds  for the true ground state energy, $E_0 \le E_X$, where the subscript $X$ 
stands for $T$ or $DMC$ as appropriate and $\Psi_{DMC}$ is the FN ground state. Within VMC, we have also access
to the variance of the energy in the trial state, $\sigma^2= \langle \Psi_T | (H-E_T)^2 | \Psi_T \rangle$.
As the variance approaches zero for any exact eigenstate, its value for a given trial wave function can be used to
quantify
the distance to the closest eigenfunction. Under the assumption that the trial energy is
closer to the ground state energy than to any of the other eigenstates, the inequality  $\sigma^2 \ge (E_0-E_T)^2$ 
leads to a lower bound  for the ground state energy \cite{Temple}:
\beq
E_0 \ge E_T - \sqrt{\sigma^2}.
\label{Temple}
\eeq
In the following, we will use the information on the variance obtained by VMC to extrapolate to the exact ground state energy.

Let us first analyse in more detail how the trial wave function approaches the ground state wave function.
Expanding our trial wave function in the exact eigenstates, $|E_j\rangle$, of energy $E_j$, we have
$|\Psi_T \rangle = \sum_j c_j | E_j\rangle$
where $c_j$ are the expansion coefficients, with $\sum_j |c_j|^2=1$ assuming normalized states.
We can now write
\beq
E_T = E_0 + \Delta_T  C_T
\label{ET_C}
\eeq
with  $C_T= \sum_{j\ne 0}^M |c_j|^2$ and $\Delta_T \equiv \sum_i (E_i-E_0) c_i^2/C_T\ge \Delta$,
where $\Delta \equiv E_1-E_0$ denotes the energy gap between ground and  first excited state of the system. 
Similarly, we obtain for the variance
\beq
\sigma^2=\overline{\Delta^2}_T C_T - \left(\Delta_T C_T \right)^2
\label{sigmaT_C}
\eeq
where $\overline{\Delta^2}_T \equiv \sum_i (E_i-E_0)^2 c_i^2/C_T \ge \Delta^2_T$.

Using Eqs.~(\ref{ET_C}) and (\ref{sigmaT_C}) we have
\beq
E_T -\sqrt{\sigma^2} =E_0 - \Delta_T C_T
\left[ \sqrt{ \frac{\overline{\Delta^2}_T}{\Delta_T^2 C_T} -1 } -1 \right],
\eeq
and we see that the expression for the lower bound, Eq.~(\ref{Temple}), 
remains valid for $C_T  \le \overline{\Delta^2}_T/2\Delta_T^2$,
or $E_T-E_0 \le \overline{\Delta^2}_T/2\Delta_T$. Note that this condition is less stringent
than the assumption that $E_T$ is closer to the ground state energy than to any of the other eigenstates used previously.

\begin{figure}
\includegraphics[width=0.5\textwidth]{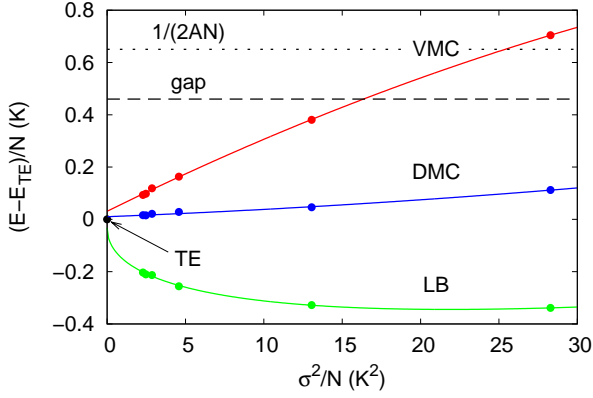}
  \caption{The VMC and DMC energies per particle of $N=26$ unpolarized $^3$He atoms in two dimensions as a 
  function of the variance $\sigma^2/N$. Each point corresponds to a different trial function (PW and 
  BF$^{(\alpha)}$ with $\alpha=0$ to 4, from higher to lower variance). The TE energy has been subtracted. 
  Their dependence is nearly linear, and their extrapolations to zero variance, the entries VMC$_{ext}$ and DMC$_{ext}$ of Table \ref{energies},
  are very close to the exact result. We further show the energy lower bound $(E_T-\sqrt{\sigma^2})/N$, whose extrapolation 
  to zero variance, the entry LB$_{ext}$ of Table \ref{energies}, is also very close to the exact result.
  The dashed line is a rough estimate of the first excited state (the difference between the two slowest 
  exponential decay constants in the fermionic signal of the TE procedure\cite{Carleo}). It shows that the 
  condition for the validity of the energy lower bound (see text) are met by the iterated backflow trial functions.
  The dotted line is the alternate estimate $1/(2AN)$ of Eq.~(\ref{a}) for the validity of the lower bound. 
  }
\label{unpolarized26}
\end{figure}

\begin{figure}
\includegraphics[width=0.5\textwidth]{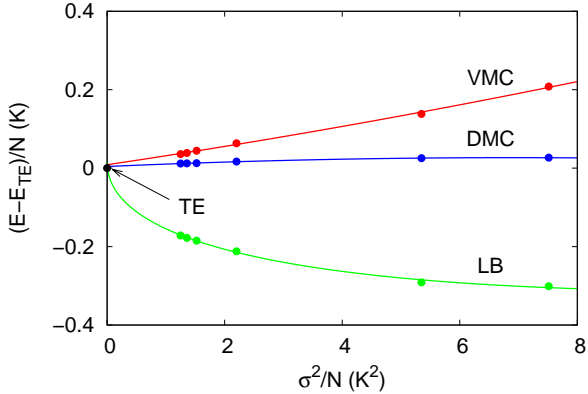}
  \caption{Same as Fig. \ref{unpolarized26}, for $N=29$ spin-polarized $^3$He atoms. 
  Both the estimate of the first excited state, $\sim 0.6$ K, and the value of $1/(2AN)$,  0.772~K, are off scale.
}
\label{polarized29}
\end{figure}

\begin{figure}
\includegraphics[width=0.5\textwidth]{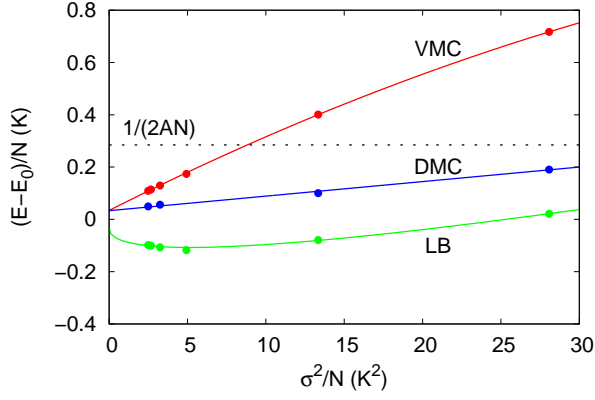}
  \caption{Same as Fig. \ref{unpolarized26}, for $N=58$ unpolarized $^3$He atoms. 
  In the lack of TE results, we take a reference energy $E_0/N$ halfway between
  VMC$_{ext}$ and LB$_{ext}$. Also, we do not have an estimate of the first excited
  state. Comparison of the dotted line with that of Fig.~\ref{unpolarized26} shows 
  that the range of $E_T-E_0$ where the lower bound is expected to be valid shrinks as $1/N$.
}
\label{unpolarized58}
\end{figure}

To go further, let us assume that the trial wave function has a significant overlap only with the ground state wave function,
whereas the components of excited staes, $c_i$ with $i > 0$, are broadly distributed. We expect this assumption to be reasonably satisfied for extended systems, where the excited states approach a continuum in the thermodynamic limit.
Improving the wave function via our iterative renormalization, the excited state contributions decrease almost uniformly, 
such that $C_T \to 0$ whereas $\Delta_T$ and $\overline{\Delta^2}_T$ remain roughly constant.
In this case we can neglect terms of order $C_T^2$ in Eq.~(\ref{sigmaT_C}) and insert it in Eq.~(\ref{ET_C}), and obtain
\beq
E_T = E_0 +  A \, \sigma^2,  \quad \text{for $\sigma^2 \to 0$}
\label{Evssigma}
\eeq
with $A = \Delta_T/\overline{\Delta^2}_T$.
Therefore, with good enough trial functions, we expect that a (nearly) linear extrapolation of the variational energy
to zero variance closely approaches the exact ground state energy, with the coefficient of the linear term
providing a numerical estimate of the validity of the lower bound of Eq.~(\ref{Temple}), i.e. 
\beq
E_T-E_0  \le 1/(2A).
\label{a}
\eeq
The lower bound, in turn, can be made stricter by extrapolation to zero variance of $E_T-\sqrt{\sigma^2}$ with a leading
square-root term.
These variance extrapolations are shown in Figs~(\ref{unpolarized26}), (\ref{polarized29}) and (\ref{unpolarized58})
and listed in Table  \ref{energies}. 

\begin{figure}
\includegraphics[width=0.5\textwidth]{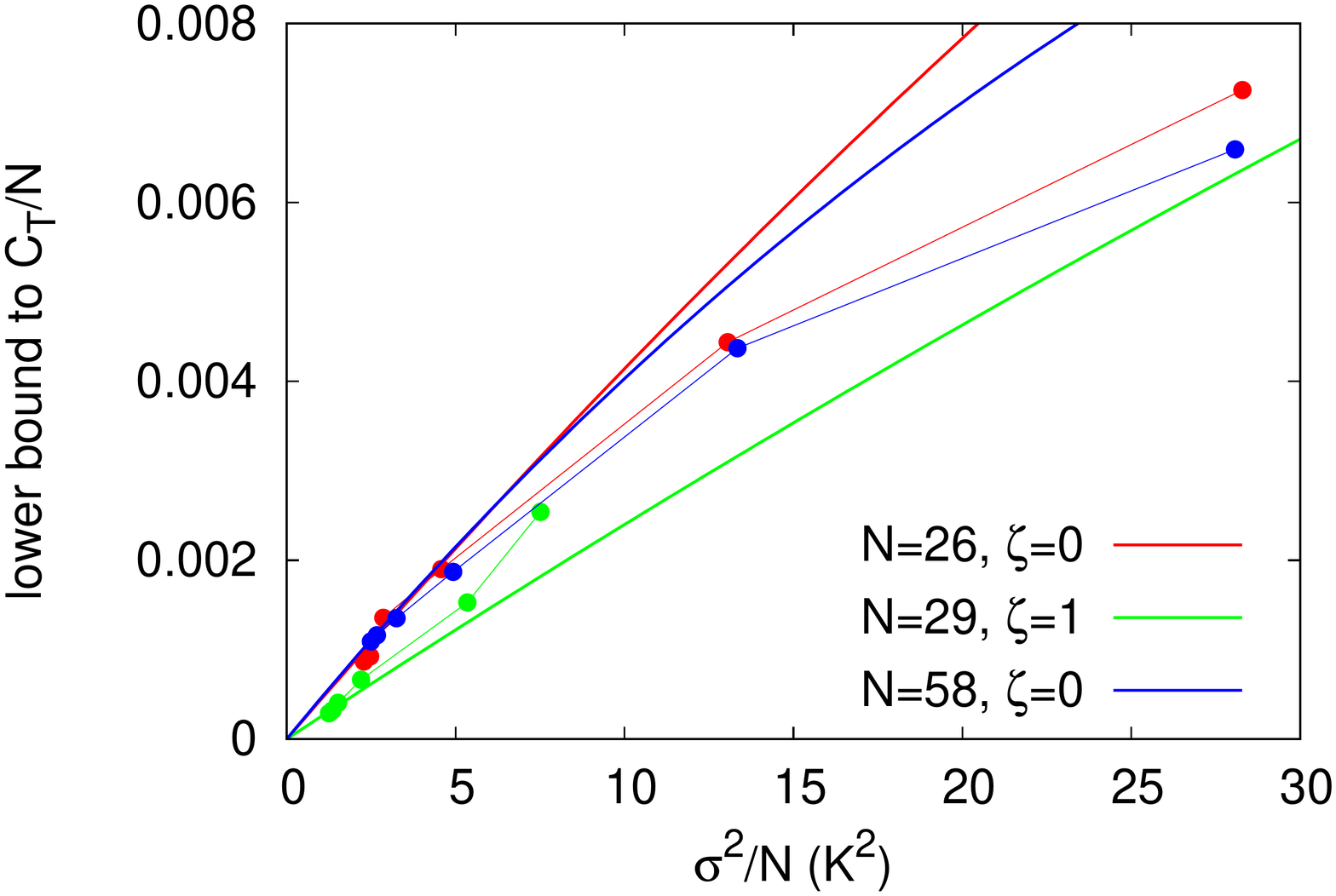}
  \caption{
Points connected by thin lines : the lower bound $\{1- \exp[-(E_T-E_0)^2/2\sigma^2]\}/N$ to the missing
overlap per particle between trial function and ground state, $C_T/N$ (with $E_0$ replaced by the zero-variance
extrapolation of $E_T$). The points refer to
trial functions from PW to BF$^{(4)}$, in order of decreasing variance. Thick lines: the approximation
$\{1-\exp[-A^2 \sigma^2]\}/N$ obtained using only the linear term of the fit of energy vs. variance.
}
\label{C_T}
\end{figure}

All of the above extrapolations are valid for $C_T \to 0$. Using a general 
estimate  for the overlap of the trial wave function with the ground state \cite{xavier}, 
$C_T \equiv 1-c_0^2 \ge 1- \exp[-(E_T-E_0)^2/2\sigma^2] \approx 1-\exp[-A^2 \sigma^2]$,
we can a-posteriori check the consistency of the energy versus variance extrapolation, see
Fig.~\ref{C_T}. 

Finally, one would like to use variance extrapolation with the fixed-node energies to obtain even better results. 
However, within DMC, the variance 
$\langle \Psi_{DMC} | (H-E_{DMC})^2 | \Psi_{DMC} \rangle$ is zero inside any nodal pocket \cite{DMCVariance} and cannot be used anymore  as a measure of the quality of the wave function.
The most natural assumption is then to postulate 
that the variance $\sigma^2$ calculated in VMC is a good 
measure of the quality of the wave function in DMC as well.
This allows us to use the same
extrapolation for DMC energies as in the case of VMC, as shown in Figs.~(\ref{unpolarized26}), (\ref{polarized29}) 
and (\ref{unpolarized58}), but without obtaining a lower bound. 
The DMC energy extrapolated to zero variance, listed in Table \ref{energies}, happens to differ
from the TE value by just the combined error bar, 10 mK for $\zeta=0$ and 4 mK for $\zeta=1$.

\section{Conclusions}

In this paper we have introduced new, highly correlated wave functions 
for accurate descriptions of normal Fermi liquids based on generalized backflow coordinates which
are iteratively improved. 
For liquid $^3$He at freezing density, the energy gain of these wave functions at the 4th iteration 
compared to the usual backflow trial wave function (0 iterations)
is about $290$mK  within VMC and $30$mK for FN-DMC.
More important, we have shown that the true ground state energy can be obtained by variance extrapolation with
intrinsic a-posteriori checks of the consistency and validity of the extrapolation. For small number of atoms, $N\sim 26$, we have shown that the
obtained results are in agreement with unbiased calculations using transient estimates,
but variance extrapolation can be used also to quantify  the fixed-node error of larger systems. 
For systems with $N=58$ atoms, the fixed-node error of our best wave function is around $20$mK.

Thus, 
apart from significant VMC and FN-DMC energy gains, 
the iterative backflow renormalization procedure also
leads to a general strategy to quantify the fixed-node error of the calculations.
In combination with finite-size extrapolations based on the analytical informations 
contained in the trial wave function \cite{FSE,FLP}, the methods presented in the paper
provide
an important step towards the control of the accuracy of QMC calculations suffering from
a Fermion sign problem.

\begin{acknowledgements}
 MH thanks Bernard Bernu and David Ceperley for discussions.
Computer time at CNRS-IDRIS is acknowledged, Project No. i2014051801. 
\end{acknowledgements}

\appendix
\section{Computational details}
Let us suppose that $q_i^\alpha$, are backflow coordinates  $i=1,\dots N$
and $\alpha=1,\dots d$, where $d$ is the spatial dimension, and 
we have already computed the
following partial derivatives
\bea
Q_{ij}^{\alpha \beta} & \equiv& \nabla_i^\alpha q_j^\beta
\label{gradQ}
\\
\widetilde{Q}_j^\beta & \equiv & \Delta q_j^\beta \equiv
\sum_{i\alpha} \nabla_i^\alpha \nabla_i^\alpha q_j^\beta
\label{delQ}
\eea 
We will further need
\beq
\overline{Q}_{lm}^{\beta \gamma} =\sum_{i \alpha} Q_{il}^{\alpha \beta} Q_{im}^{\alpha \gamma}
\label{Fmat}
\eeq
which is already needed for computation of the local energy of the Slater determinant using
orbitals based on the above backflow coordinates \cite{Kwon93}, which we will shortly remind.

\subsection{Backflow determinant}
The gradient and the laplacian of a determinant, $D= \det
\varphi_{ki}$, with backflow coordinates
in the orbitals, $\varphi_{ki} \equiv \varphi_k(\qvec_i)$, can be calculated as follows
\bea
\nabla_i^\alpha \log D &=& \sum_{j \beta} F_{jj}^\beta Q_{ij}^{\alpha \beta}
\label{gradD}
\\
\Delta \log D &=&
\sum_{i \alpha} F_{ii}^\alpha \widetilde{Q}_i^\alpha
+ \sum_{i \alpha \beta} \left[ \sum_m V_{im} \varphi_{mi}^{\alpha \beta} \right] \overline{Q}_{ii}^{\alpha \beta}
\nonumber
\\
&&- \sum_{ij \alpha \beta} F_{ij}^\alpha F_{ji}^\beta \overline{Q}_{ji}^{\alpha \beta}
\label{delD}
\eea
where 
\beq
\varphi_{ki}^\alpha \equiv \frac{\partial \varphi_{ki}}{\partial q_i^\alpha}, \quad
\varphi_{ki}^{\alpha \beta}  \equiv \frac{\partial^2 \varphi_{ki}}{\partial q_i^\alpha \partial q_i^\beta},
\quad
F_{ij}^\alpha = \sum_k V_{ik} \varphi_{kj}^\alpha
\label{F}
\eeq
and $V_{ik}$ is the inverse of the backflow matrix
\beq
V_{ik}= \frac{1}{D} \frac{\partial D}{\partial \varphi_{ki}},Ê\quad
\sum_k V_{ik} \varphi_{kj} = \delta_{ij}
\label{inverse}
\eeq
The computational complexity is of order $N^3$ for the inversion of the orbital matrix, $V_{ik}$, as
well as for the computation of the matrices $F_{ij}^\alpha$ and $\overline{Q}_{ij}^{\alpha \beta}$.

Note that this part of the calculations does not depend on the specific form of the backflow coordinates.
The computation of the gradient and laplacian of the local energy based on Eq~(\ref{gradD}) and
Eq~(\ref{delD})
do only depend on the actual values of the orbital matrix, $\varphi_{ki}$, and its partial derivatives, Eqs.~(\ref{F}),
and on the gradient and laplacian  of the backflow coordinates, Eq~(\ref{gradQ}) and Eq.~(\ref{delQ}). 
Therefore, Eq.~(\ref{gradD}) and Eq.~(\ref{delD})  can be still used to calculate the local energy of determinants
containing  iterated backflow coordinates, as long as their derivatives are provided in the form of Eqs~(\ref{gradQ}),
(\ref{delQ}), and (\ref{Fmat}).

\subsection{Iterated Jastrow correlations}
We can now build a Jastrow factor based on the distances between two quasi-particles,
\beq
U=\sum_{l<m} u(q_{lm})
\eeq
where $u$ denotes the function and  $u'$  ($u''$) its first  (second) derivative.
The gradient of the Jastrow factor can then be calculated by the chain rule
\beq
\nabla_i^\alpha U= \sum_{l \beta} V_l^\beta Q_{il}^{\alpha \beta}, \quad
V_l^\beta = \sum_{m\ne l} \frac{u'(q_{lm})}{q_{lm}} q_{lm}^\beta
\eeq
and
\beq
\Delta U
= \sum_{l \beta} V_l^\beta \widetilde{Q}_l^\beta
+ \sum_{l \ne m} \sum_{\beta \gamma} W_{lm}^{\beta \gamma} \left[ \overline{Q}_{ll}^{\beta \gamma} - \overline{Q}_{lm}^{\beta \gamma}
\right]
\eeq
with
\beq
W_{lm}^{\beta \gamma}=\left( u''(q_{lm}) - \frac{u'(q_{lm})}{q_{lm}} \right) \frac{q_{lm}^\beta q_{lm}^\gamma}{q_{lm}^2}
+ \delta_{\beta \gamma} \frac{u'(q_{lm})}{q_{lm}}
\eeq
We see that the overall cost of the quasi-particle Jastrow factor and its derivatives needed for the local energy 
is of order of $N^3$, needed to build the matrix $\overline{Q}_{lm}^{\beta \gamma}$,
Eq.~(\ref{Fmat}). Since this matrix  is already needed 
in the calculation of usual backflow wave function \cite{Kwon93}, the iterated
Jastrow does not lead to a  significant slow down compared to the usual backflow.

\subsection{Iterated backflow coordinates}

We now construct new backflow coordinates
\beq
y_i^\alpha=  \sum_{j \ne i} q_{ij}^\alpha \eta(q_{ij})
\eeq
where $\eta$ is the corresponding potential. In order to calculate the local energy for backflow orbitals in the Slater
determinant based on $y_i^\alpha$, we need the following derivatives
\beq
Y_{ij}^{\alpha \beta} \equiv \nabla_i^\alpha y_j^\beta, \quad
\widetilde{Y}_i^\alpha \equiv \Delta y_i^\alpha
\label{Yderv}
\eeq
In order to calculate them, we will use the chain rule, based on the following partial derivatives
\bea
\frac{\partial y_j^\beta}{\partial q_i^\alpha}
&=&\delta_{ij} \sum_n \dot{y}_{in}^{\alpha \beta} - \dot{y}_{ij}^{\alpha \beta} \nonumber \\
\frac{\partial^2 y_k^\gamma}{\partial q_i^\alpha \partial q_j^\beta}
&=&
\delta_{ijk} \sum_n \ddot{y}_{kn}^{\alpha \beta \gamma}
-\delta_{jk} \ddot{y}_{ki}^{\alpha \beta \gamma}
- \delta_{ij} \ddot{y}_{jk}^{\alpha \beta \gamma}
- \delta_{ik} \ddot{y}_{kj}^{\alpha \beta \gamma}
\nonumber
\eea
where
\bea
\dot{y}_{ij}^{\alpha \beta} &=&
 \frac{\eta'(q_{ij})}{q_{ij}} q_{ij}^\alpha q_{ij}^\beta + \eta(q_{ij}) \delta_{\alpha \beta}
 \nonumber
 \\
 \ddot{y}_{ij}^{\alpha \beta \gamma} &=&
 \left[ \eta''(q_{ij} - \frac{\eta'(q_{ij})}{q_{ij}} \right] \frac{q_{ij}^\alpha q_{ij}^\beta q_{ij}^\gamma}{q_{ij}^2}
 \nonumber
 \\
 && + \frac{\eta'(q_{ij})}{q_{ij}} \left[ q_{ij}^\alpha \delta_{\beta \gamma}
 + q_{ij}^\beta \delta_{\alpha \gamma} + q_{ij}^\gamma \delta_{\alpha \beta} \right]
 \nonumber
\eea
and we have used that $\dot{y}_{ii}^{\alpha \beta} = \ddot{y}_{ii}^{\alpha \beta \gamma} =0$.

The final derivatives needed, Eqs.~(\ref{Yderv}), can then be written as
\bea
Y_{ij}^{\alpha \beta} &=& \sum_{n \gamma} \dot{y}_{jn}^{\gamma \beta} 
\left[ Q_{ij}^{\alpha \gamma} - Q_{in}^{\alpha \gamma} \right]  \nonumber
\\
\widetilde{Y}_i^\alpha &=&
\sum_{n \beta} \dot{y}_{in}^{\alpha \beta} \left[ \widetilde{Q}_i^\beta - \widetilde{Q}_n^\beta \right]
\nonumber
\\
&& + \sum_{n \alpha \beta} \ddot{y}_{in}^{\alpha \beta \gamma}
\left[ \overline{Q}_{ii}^{\beta \gamma}
+ \overline{Q}_{nn}^{\beta \gamma}
-  2 \overline{Q}_{in}^{\beta \gamma}
\right]
\eea
Again, these operations can be done in order of $N^3$ computations.

\subsection{Iterated $n$-body correlations}

Above we have explicitly shown how to calculate gradient and laplacian of a scalar two-body Jastrow potential
and of quasi-particle coordinates constructed from backflow coordinates. 
The structure of our three-body correlation in Eq.~(\ref{Jastrowrenorm}) is actually a scalar product between 
two vectors with identical structure as the backflow coordinates. Gradient and laplacian of the three
body term can therefore be calculated from those of the vectors using the chain rule. 
Generalizations to build iterated many-body
Jastrow and backflow coordinates based on quasi-particle tensors \cite{BF3} are straightforward and do not
increase the complexity of the calculation.   

\subsection{Higher order iterations}

At the zeroth order of iteration, the backflow coordinates,
$\qvec_i$, are symmetric functions of the bare coordinates. Higher order iterations of the backflow
are built from symmetric expressions based on 
the previous backflow coordinates, such that the overall wave function remains antisymmetric.
Above, we have explicitly shown how to calculate the gradient and the laplacian of the quasi-particle coordinates of the
first backflow iteration, $\qvec_i^{(1)} \equiv \qvec_i+ \yvec_i$ without increasing the overall complexity of the calculation.
These are the only additional information needed to calculate the local energy of the first iterated backflow determinant,
and from the structure it is clear that this procedure can be iterated to higher order without increasing the
complexity of the calculations.

\newpage

\begin{thebibliography}{99}
\bibitem{DMCVariance} Unless the nodes are exact, the FN-DMC wave function will have a discontinuous derivative
at the nodal surface. Nevertheless, one can show that the energy expectation value of a FN-DMC wave function still provides an upper bound to the true ground state wave function having continuous derivatives everywhere.
\end{thebibliography}
\end{document}